\documentclass{article}
\usepackage{spconf,amsmath,graphicx}
\usepackage{multirow}
\usepackage{hyperref}
\usepackage{amssymb}
\hypersetup{
    colorlinks=true,
    linkcolor=red,
    filecolor=magenta,      
    urlcolor=red,
}

\newcommand{\newpara}[1]{\vspace{8pt}\noindent\textbf{#1}}

\title{Exploring Auditory Acoustic Features for the Diagnosis of COVID-19}

\name{Madhu R. Kamble, Jose Patino, Maria A. Zuluaga and Massimiliano Todisco\thanks{The first author is supported by the RESPECT project funded by the French Agence Nationale de la Recherche (ANR).}}
\address{EURECOM, Sophia Antipolis, France}

\begin{document}

\maketitle
\begin{abstract}
The current outbreak of a coronavirus, has quickly escalated to become a serious global problem that has now been declared a Public Health Emergency of International Concern by the World Health Organization. Infectious diseases know no borders, so when it comes to controlling outbreaks, timing is absolutely essential. It is so important to detect threats as early as possible, before they spread. After a first successful DiCOVA challenge, the organisers released second DiCOVA  challenge with the aim of diagnosing COVID-19 through the use of breath, cough and speech audio samples. This work presents the details of the automatic system for COVID-19 detection using breath, cough and speech recordings. We developed different front-end auditory acoustic features along with a bidirectional Long Short-Term Memory (bi-LSTM) as classifier. The results are promising and have demonstrated the high complementary behaviour among the auditory acoustic features in the Breathing, Cough and Speech tracks giving an  AUC of 86.60\% on the test set.
\end{abstract}

\section{Introduction}
\label{sec:intro}

Coronavirus disease, so-called COVID-19, is an infectious disease caused by the recently discovered coronavirus, the SARS-CoV-2. This disease has spread rapidly worldwide over the past year, causing a global crisis with serious health, social and economic consequences. To put an end to this pandemic, various initiatives are being carried out worldwide, including the development of new systems for rapid diagnosis of the disease.

Recently, the DiCOVA 2021 Challenge~\cite{muguli21_interspeech} was carried out to promote research in development of systems for the detection of COVID-19 through recordings of respiratory sounds.
Several systems have been proposed to detect the COVID-19 signature within acoustic indicators~\cite{kamble21_interspeech,das21_interspeech,bhosale21_interspeech,avila21_interspeech,ritwik21_interspeech,karas21_interspeech}. Only a few of them focused on the study of acoustic clues, giving more emphasis to classifiers. The study reported in~\cite{harvill21_interspeech} explores the Autoregressive Predictive Coding (APC) to pre-train a unidirectional LSTM and spectral augmentation. In~\cite{sodergren21_interspeech},  authors used ComParE 2016 feature set, and two classical machine learning models, namely Random Forests, and Support Vector Machines (SVMs). The use of breathing patterns for the diagnosis of COVID-19 is studied in~\cite{deshpande21_interspeech}. COVID-19 detection by means of a Contextual Attention Convolutional Neural Networks and gender information is studied in~\cite{mallolragolta21_interspeech}.

The first COVID-19 challenge consisted on 2 tracks: Track-1 focused on diagnosing COVID-19 using cough sounds, while Track-2 focused on a collection of breath, sustained vowel phonation, and a number of counting speech recordings. As a follow up of the first successful DiCOVA 2021 challenge, the second DiCOVA challenge has been organised~\cite{Dicova2_Baseline}. The second challenge aimed at 4 different tracks, namely, breathing, cough, speech and fusion. The organisers provided a baseline system for the second challenge based on Log Mel Spectrogram front-end and Bidirectrional Long Short-Term Memory (bi-LSTM) back-end.

In this paper, we describe and propose a system for automatic COVID-19 detection presented on all the four different tracks of the second DiCOVA challenge. Our system focuses more on features than classifiers by using 4 perceptually-motivated acoustic features at front-end. The  features we explored are Teager energy operator cepstral coefficients (TECCs), Instantaneous Amplitude Cepstral Coefficients (IACCs), Constant Q-Cepstral Coefficients (CQCCs) and Filterbank Constant Q Transform (FBCQT)~\cite{kamble2019analysis,kamble2018effectiveness,todisco2017constant}, along with the bi-LSTM classifier.

The remainder of this paper is organized as follows. Section~\ref{sec:architecture} presents the technical details of auditory acoustics features used for the detection of COVID-19. Section~\ref{sec:dicova} describes the second DiCOVA challenge database. Experimental setup and results are presented in Section~\ref{sec:expt_setup} and Section~\ref{sec:expt_results}, respectively. Finally, the main conclusions of this work and future research lines are drawn in Section~\ref{ssec:conclusion}.
\vspace{-.4cm}
\section{Auditory Acoustics features}

\label{sec:architecture}
In this section we discuss the acoustic features used to diagnose COVID-19 from breath, cough and speech.

\vspace{-.2cm}
\subsection{MelSPEC}

Studies have shown that humans do not perceive frequencies on a linear scale. We are better at detecting differences in lower frequencies than higher frequencies~\cite{davis_mfcc}.
The Mel spectrum contains a short-time Fourier transform (STFT) for each frame of the spectrum (energy/amplitude spectrum), from the linear frequency scale to the logarithmic Mel-scale, and then goes through the filter bank to get the eigen vector, these eigenvalues can be roughly expressed as the distribution of signal energy on the Mel-scale frequency.

\subsection{TECC and ESA-IACC}
\begin{figure}[h]
		\centering
		\includegraphics[width=\linewidth]{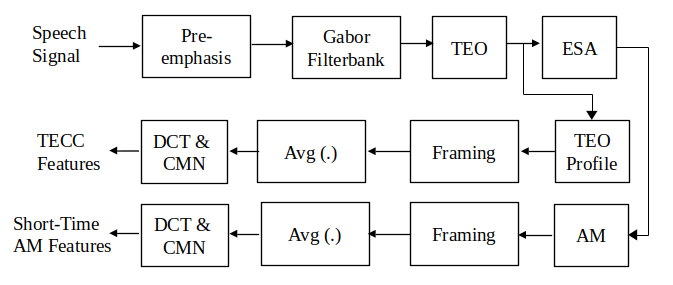}\\ 
			\vspace{-0.4 cm}
		\caption{Block diagram of TECC, and ESA-IACC feature sets.}
		\label{fig:Block_diagram}
		 	\vspace{-0.6 cm}
	\end{figure}

The Teager Energy Operator (TEO) ($\psi\{\cdot\}$) track the running estimate of instantaneous energy fluctuations of the narrow-band signal. The Teager energy profile obtained from the bandpass filter is further given to the Energy Separation Algorithm (ESA) to isolate the Instantaneous Amplitude (IA) ($a_i[n]$) and Instantaneous Frequency (IF) ($\Omega_i[n]$) and is given as \cite{maragos1992separating, maragos1993amplitude,quatieri2006discrete}:
	\begin{equation}
	\Psi_d\{x_i[n]\}=x_i^2[n]-x_i[n-1]x_i[n+1]\approx a_i^2[n]\Omega_i^2[n],
	\label{Eq: ESA-TEO}
	\end{equation}
where $x_i[n]$ is $i^{th} $ bandpass filtered signal.
	\begin{equation}
	a_i[n]\approx\frac{2\Psi_d\{x[n]\}}{\sqrt{\Psi_d\{x[n+1]-x[n-1]\}}},
	\label{Eq:AM}
	\end{equation}
	where $x_i[n]$ is $i^{th} $ bandpass filtered signal.
The block diagram of Teager Energy Cepstral Coefficients (TECC) and Energy Separation Algorithm Instantaneous Amplitude Cepstral Coefficients (ESA-IACC) feature set is shown in Figure \ref{fig:Block_diagram}. 
The TECC feature set is computed as per our earlier studies in \cite{kamble2019analysis,kamble2021CSL} and ESA-IACC feature set according to \cite{kamble2018effectiveness,kamble2020speechcommunication}. 
\vspace{-0.3 cm}

\subsection{Filterbank CQT and CQCC}
The constant Q transform (CQT) is a perceptually motivated approach to time-frequency analysis introduced by Youngberg and Boll~\cite{youngberg1978constant} in 1978 and refined over the last few decades by Brown~\cite{brown1991calculation}.
In contrast to Fourier-based approaches, the CQT gives a greater frequency resolution for lower frequencies and a greater temporal resolution for higher frequencies, which emulate the human auditory perception.
The CQT of a discrete signal $x(n)$ is defined by:
\begin{equation}
X^{CQ}(k,n)=\sum_{j=n-\left \lfloor N_k/2 \right \rfloor}^{n+\left \lfloor N_k/2 \right \rfloor}x(j)a_{k}^{*}(j-n+N_{k}/2)
\label{eq:CQT}
\end{equation}
where $k = 1, 2,..., K$ is the frequency bin index, $a_k(n)$ are the basis functions, $*$ is the complex conjugate and $N_k$ is a variable window length. The center frequencies $f_{k}$ are defined according to
$f_k=2^{(k-1)/(B)}f_1$,
where $f_{k}$ is the center frequency of bin $k$, $f_{1}$ is the center frequency of the lowest frequency bin and $B$ is the number of bins per octave. 

The filter selectivity $Q$ which reflects the ratio between  center frequency and  bandwidth is constant and defined as:
\begin{equation}
Q=\frac{f_k}{f_{k+1}-f_k}=(2^{1/B}-1)^{-1}
\end{equation}
In practice, $B$ determines time-frequency resolution trade-off.

\subsubsection{FBCQT}
Similar to MelSPEC, FBCQT is calculated  filtering $X^{CQ}(k,n)$ with a filterbank composed of $n_{fb}$ triangular filters equally spaced along the linear-scale, and then calculating the logarithm of the energy in each band.

\subsubsection{CQCC}
Constant Q cepstral coefficients (CQCCs) were introduced recently and successfully in the context of fake audio detection~\cite{todisco2017constant}. 
CQCC features are based on a combination of the constant Q transformation (CQT) and cepstral analysis.
The cepstral coefficients are calculated from the transformation at constant Q, which imply a re-sampling in frequency domain from the geometric scale to a linear scale, according to:
\begin{equation}
CQCC(p)=\sum_{l=0}^{L-1}\mathrm{log}\left  |\bar{X}^{CQ}(l)  \right |^{2} \mathrm{cos}\left [ \frac{p\left ( l-\frac{1}{2}  \right )\pi}{L} \right ]
\label{eq:cqcc}
\end{equation}
where $\bar{X}^{CQ}$ is the linearised CQT-derived spectrum, $l$ is the linear-scale index and $p = 0...L-1$.  
The full CQCC extraction algorithm is described in~\cite{todisco2017constant}.

\section{Second DiCOVA Challenge Dataset}
\label{sec:dicova}
After the first successful DiCOVA Challenge, the organisers released the second DiCOVA challenge focusing on four different tracks, namely, breathing, cough, speech, and fusion~\cite{Dicova2_Baseline}.

The training/validation set for the all the tracks contains 965 audio files stored in .FLAC format at 44.1 kHz sampling frequency. Each audio file corresponds to a single subject. This set comprises an audio recording from 172 COVID-19 positive subjects and 793 COVID-19 negative subjects. Gender and age of the subjects is also provided as extra metadata. Validation set are performed using 5-fold cross validation  from train lists. The test set consists of 471 audio files with the same format as the training/validation set, but with the COVID-19 status hidden from the participants. The detailed information of the respective tracks are reported below:
\newpara{{Track 1: Breathing}} - The goal of this track is to use the key differences and analyze the breathing signal from COVID-19 positive and negative subjects that can contribute towards the detection of the disease. In total, the samples provided by the organizers include the data of 965 subjects that is further split into train and validation set. The dataset contains lists corresponding to a 5-fold cross validation split.

\newpara{{Track 2: Cough}} - The goal of this track is to use cough sound recordings from COVID-19 positive and negative subjects. The validation set is composed of cough audio data from 965 subjects. The dataset also contains lists corresponding to a 5-fold cross validation split. 

\newpara{{Track 3: Speech}} - Similar to the Track 1, the goal of this track also aims to detect the COVID-19 disease using the speech signals from positive and negative COVID-19 subjects. The data distribution and 5-fold cross validation is also similar to the previous tracks.

\newpara{{Track 4: Fusion}} - In this track, all the scores from the breath, speech and cough track of the corresponding folds are used to have the simple arithmetic mean of those particular folds. The validations scores are the concatenation of all the folds.

\vspace{-0.3 cm}
\subsection{Baseline and evaluation metrics}
The organisers of the challenge provide a baseline system for 4 different tracks based on log Mel spectrogram. The back-end classifier used is a bidirectional Long Short-Term Memory (bi-LSTM)
Classification performance evaluation is measured using traditional detection metrics, namely, true positive rate (TPR) and false positive rate (FPR) over a range of decision thresholds. From these metrics, the probability scores for each audio file are used to compute the receiver operating characteristic (ROC) curve, and the area under the curve (AUC) metric to quantify the model performance~\cite{FAWCETT2006ROC}.

\begin{table*}
\centering
\caption{Results on validation set for  Breathing and Cough on Second DiCOVA  Challenge database in terms of AUC (\%).}

\begin{tabular}{|c|c|c|c|c|c|c|c|c|c|c|c|c|c|} 
\hline
\multirow{2}{*}{Folds} & \multicolumn{5}{c|}{\textbf{Breathing}}  & \multicolumn{5}{c|}{\textbf{Cough}}       \\ 
\cline{2-11}
        & MelSPEC & TECC & IACC &  CQCC& FBCQT & MelSPEC & TECC & IACC &  CQCC & FBCQT \\ 
\hline
0 &76.40&74.40&78.50 &78.60&80.00&66.80 & 74.30&76.00&79.40 & 79.60 \\ 
\hline
1  &75.00&73.90& 80.30&78.10&80.00&77.10& 67.50&78.70&74.50 & 84.20   \\  
\hline
2  &75.00&74.80& 76.60&78.30&74.80&77.40& 69.80&74.60& 77.70& 78.80  \\ 
\hline
3   &80.30&72.30& 77.70&75.00&78.90&74.80&81.00 &74.70&78.20 &  77.50\\ 
\hline
4   &82.10&87.80& 86.40&82.60&88.90&77.40& 73.40&79.00& 83.20& 77.90  \\ 
\hline
Avg&77.76&76.64&79.90&78.52&\textbf{80.52}&74.70&73.20&76.60&78.60& \textbf{79.60}  \\\hline
\end{tabular}

\label{Tab:valid_results}
\end{table*}

\begin{table*}
\centering
\caption{Results on validation set for Speech and Fusion on Second DiCOVA  Challenge database in terms of AUC (\%).}

\begin{tabular}{|c|c|c|c|c|c|c|c|c|c|c|c|c|c|} 
\hline
\multirow{2}{*}{Folds}  & \multicolumn{5}{c|}{\textbf{Speech}}     & \multicolumn{5}{c|}{\textbf{Track Fusion}}      \\ 
\cline{2-11}
        & MelSPEC & TECC & IACC &  CQCC& FBCQT & MelSPEC & TECC & IACC &  CQCC & FBCQT \\ 
\hline
0 &74.60 &70.70 &75.10&74.20& 78.10  &75.00&77.30 &79.90&80.70&   82.20\\ 
\hline
1 &85.90& 79.20&80.30&82.50& 87.00 &82.90& 76.90&83.90&80.30&  86.70 \\  
\hline
2   &81.20 &75.90 &80.60&75.50&  76.50 &82.70& 78.10&80.70&80.10&  79.90 \\ 
\hline
3  &78.90 & 76.80&81.80&76.80&  77.70&81.20&81.90 &82.20&78.90&  81.60 \\ 
\hline
4   &83.80 & 81.00&87.40&81.70& 85.90  &88.40&87.40 &90.90&88.00&90.50\\ 
\hline
Avg &80.88&76.72&\textbf{81.04}&78.14& \textbf{81.04}  &82.04&80.20&83.52&81.60&  \textbf{84.18} \\\hline
\end{tabular}

\label{Tab:valid_results2}
\end{table*}

\section{Experimental Setup}

\label{sec:expt_setup}
We have performed the experiments on the second DiCOVA Challenge database.

Five acoustic features discussed in Section~\ref{sec:architecture} have been used along with a cascade of two bi-directional long-short term memory (bi-LSTM) and a fully connected neural network with an encoder-decoder style network.  The encoder consists of two bi-LSTM layers with 128 units in both the forward and back-ward direction.   
This is fully connected neural network comprising of 256 nodes in the first layer and 64 nodes and a tanh(·) non-linearity in the second layer.  Finally, a single node output,  passed  through  a  sigmoid  non-linearity  is  obtained  as  the COVID-19 probability score for the input feature matrix.

Parameters used to extract acoustic features are detailed hereafter.\\
\noindent\textbf{MelSPEC}: The MelSPEC feature set was extracted, similarly for the baseline system, using 64-dimensional log Mel spectrogram with $\Delta$ and $\Delta\Delta$ resulting in total 192-D feature vector. \\

\noindent\textbf{TECC}:	The TECC feature set was extracted using 40 Mel-spaced Gabor filterbank with $f_{min}$=10 Hz, and $f_{max}$=$fs$/2 Hz \cite{kamble2019analysis}. For each subband filtered signals, we obtain 40-D static features appended along with their $\Delta$ and $\Delta\Delta$ coefficients resulting in 120-D feature vector.\\
 
\noindent\textbf{ESA-IACC}:	The ESA-IFCC feature set was extracted using same parameters as used for TECC feature set expect the frequency scale in Gabor filterbank, here we used linearly-spaced Gabor filterbank. However, ESA-IACC feature set is computed with the pre-processing technique and cepstral mean normalization (CMN) technique for COVID-19 classification task.\\

\noindent\textbf{CQCC}:  The CQCC features are extracted with a maximum frequency of $F_{max} = F_{NYQ}$, where $F_{NYQ}$ is the Nyquist frequency of 44.1kHz.   
The minimum frequency is set to $F_{min} = F_{max}/2^9 \simeq 43$Hz (9 being the number of octaves). 
The number of bins per octave $B$ is set to 96. Only 20 static coefficients (with log-energy) were considered, resulting in total 60-dimensional (D) feature vector (including \textit{20}-$\Delta$ and \textit{20}-$\Delta\Delta$).\\

\noindent\textbf{FBCQT}:  The Filterbank CQT features set were extracted using 63-dimensional log linearised CQT with with a maximum frequency of $F_{max} = F_{NYQ}/2$ and a minimum frequency of $F_{min} = F_{max}/2^10 \simeq 43$Hz, with $\Delta$ and $\Delta\Delta$ resulting in total 189-D feature vector. \\   

	\vspace{-.5 cm}
\section{Experimental Results}
\label{sec:expt_results}
The results in terms of AUCs obtained on the validation folds for Breathing and Cough tracks are reported in Table \ref{Tab:valid_results} and for Speech and Fusion tracks reported in Table \ref{Tab:valid_results2}.  For each fold the classifier is trained using the training data and evaluated on the validation data. The average validation AUC denotes the average over the AUCs for the 5 folds. 
The acoustic features considered have their strengths and weaknesses and therefore the AUC for some folds and tracks are better compared to other folds and tracks. For all the tracks of validation set, FBCQT gave the higher AUC compared to other features. In particular, for breathing track FBCQT gave an average AUC of 80.52\%. For Cough, Speech and Track Fusion it yielded an average AUC of 79.60\%, 81.04, and 84.18\%, respectively. IACC and FBCQT feature obtained the same AUC of 81.04\% for the speech track.

We now focus on the results obtained on the blind test set that we submitted to the challenge. For evaluation on the test dataset, the COVID-19 positive likelihood score for each file was computed by taking the average over the score outputs from the 5 validation fold models. As discussed earlier on validation set, for all the tracks FBCQT gave high AUC compared to all the features considered here. However, on test set the results were contradictory to the validation set. The FBCQT feature did not perform well on test set and results in lower AUC for all the tracks compared to all the other features taken into consideration.

The best and second best system on the test set includes MelSPEC and CQCC features giving an AUC of 84.50\% and 82.21\% on breathing track and 74.89\% and 76.98\% on cough track, respectively.
On speech track, MelSPEC and IACC feature set are the best and second best giving an AUC of 84.70\% and 80.92\%, respectively, as can be viewed from Table \ref{Tab:test_results}.

\begin{table}[!h]
	\centering
	\caption{Single System results on test set of the Second DiCOVA  Challenge Database}
	\setlength\tabcolsep{3.8pt}
	\begin{tabular}{ *{7}{c}}
		\hline\hline
		 & \multicolumn{5}{c}{Single System}  \\ \cline{1-6} 
	Subset	&  MelSPEC & TECC &IACC&CQCC& FBCQT\\
		\hline  
		Breathing   &\textbf{84.50}&  67.92&  79.02&82.21&55.10\\
		Cough &74.89 &68.31	&71.55 &\textbf{76.98} &52.20 \\	
	Speech  & \textbf{84.26}&77.41 &80.92&76.90&51.90\\ 
	Fusion &\textbf{84.70}&77.75&83.26&82.01&53.00\\
 	    \hline\hline
	\end{tabular}
	\label{Tab:test_results}

\end{table}

Last but not least, we also report fusion experiments to understand the complementary information that is present in each acoustic feature under investigation.
Systems were selected each time by adding the next worst in terms of AUC according to the Track Fusion results reported in Tables~\ref{Tab:valid_results2}.
Fusion results are shown in Table~\ref{Tab:fusion}. Unexpectedly, the best combination results in MelSPEC + IACC + FBCQT feature set giving an AUC of 86.60\%. All the combinations outperform the single system based on MelSPEC except the IACC + FBCQT.

\vspace{-0.3 cm}
\begin{table}[!h]
	\centering
	\caption{System's Fusion on test set of the Second DiCOVA  Challenge Database}
	\setlength\tabcolsep{3.2pt}
	\begin{tabular}{ *{7}{c}}
		\hline\hline
		 \multicolumn{6}{c}{System's Fusion}  \\ \cline{2-7} 
		  &MelSPEC & TECC &IACC&CQCC&FBCQT&AUC\\ \hline
         &$\times$&$\times$&$\checkmark$&$\times$&$\checkmark$&82.70\\
         &$\checkmark$&$\times$&$\checkmark$&$\times$&$\checkmark$&\textbf{86.60}\\
		 &$\checkmark$&$\times$&$\checkmark$&$\checkmark$&$\checkmark$&86.00\\
		 &$\checkmark$&$\checkmark$&$\checkmark$&$\checkmark$&$\checkmark$&85.80\\
		 
		 \cline{1-4} 
		
		\hline\hline
	\end{tabular}
	\label{Tab:fusion}

\end{table}

	\vspace{-.4 cm}
\section{Summary and Conclusions}
\label{ssec:conclusion}
This paper reports on the exploration of acoustic cues using different auditory-based features for the diagnosis of COVID-19. Particularly, the systems presented are based on 5 different acoustic features based on Mel frequency scale, Teager energy operator, speech demodulation, and constant Q transform. For a proper comparison of these features we use the same back-end consisting of a bi-LSTM network.
The FBCQT system outperforms all the proposed systems for the validation set on all the tracks, however, for the test set, it demonstrated a low capacity for generalisation with limited accuracy.
Fusion experiments showed that the features considered are highly complementary. The best fusion gave an AUC of 86.60\% for the MelSPEC + IACC + FBCQT feature combination on the test set, which led us to a result between the challenge baseline system (84.70\%) and the challenge winner system (88.44\%).

\newpage

\small
\bibliographystyle{IEEEtran}
\bibliography{template}

\end{document}